\documentclass[preprint,showpacs,preprintnumbers,amsmath,amssymb]{revtex4}
\usepackage{graphicx}
\usepackage{dcolumn}
\usepackage{bm}
\usepackage{mathrsfs}

\begin{document}

\title{Spinor field realizations of the non-critical $W_{2,4}$ string
based on the linear $W_{1,2,4}$ algebra
\footnote{Supported by the National Natural Science Foundation of China, grant No. 10275030.}}

\author{Zhang Li-Jie$^1$}
\author{Liu Yu-Xiao$^2$}

\thanks{Corresponding author}\email{liuyx01@st.lzu.edu.cn}

\affiliation{$^1$Department of Mathematics and Physics, Dalian Jiaotong University, Dalian 116028, P. R. China}

\affiliation{$^2$Institute of Theoretical Physics, Lanzhou University, Lanzhou 730000, P. R. China}

\begin{abstract}
In this paper, we investigate the spinor field realizations of the
$W_{2,4}$ algebra, making use of the fact that the $W_{2,4}$
algebra can be linearized through the addition of a spin-1
current. And then the nilpotent BRST charges of the spinor
non-critical $W_{2,4}$ string were built with these realizations.
\end{abstract}


\maketitle

\section{Introduction}\label{Introduction_W2s}
As is well known, tentative extensions of string-theory based on
extra bosonic symmetry (W-symmetry) on the worldsheet are called
W-strings. The theories on $W$ algebra or $W$ string have found
remarkable applications since their discoveries
\cite{Zamolodchikov1985,Fateev1987}, they appear in $W$ gravity,
the quantum Hall effect, black holes, lattice models of
statistical mechanics at criticality, membrane theories and
other physical models \cite{Leznov1989} and so on.

The BRST charge of $W_{3}$ string was first constructed in
\cite{Thierry-Mieg1987}, and the detailed studies of it can be
found in \cite{Thierry-Mieg1987,Pope PLB 1992,LuHong1992}. From
these researches it appears that the BRST method has turned out to
be rather fruitful in the study of the critical and non-critical
$W$ string theories \cite{Becchi1974}. A natural generalization of
the $W_{3}$ string, i.e. the $W_{2,s}$ strings, is a higher-spin
string with local spin-2 and spin-$s$ symmetries on the
world-sheet. Much work has been carried out on the scalar field
realizations of $W_{2,s}$ strings \cite{LuHong1992,LuHong CQG 1994 11 939,Bershadsky PLB 1922 292 35}, and the BRST charges for such
theories have been constructed for $s=4,5,6,7$ \cite{LuHong CQG 1994 11 939}. Later in Ref. \cite{ZhaoSC PLB 2000} we pointed out the
reason that the scalar BRST charge is difficult to be generalized
to a general $W_{N}$ string. At the same time, we found the
methods to construct the spinor field realizations of $W_{2,s}$
strings and $W_{N}$ strings \cite{ZhaoSC PLB 2000,ZhaoSC PLB
2001}. Subsequently, we studied the exact spinor field
realizations of $W_{2,s}(s=3,4,5,6)$ strings and $W_{N}(N=4,5,6)$
strings \cite{ZhaoSC PLB 2000,ZhaoSC PLB 2001,ZhaoSC PRD 2001}.
Recently, the authors have constructed the nilpotent BRST charges
of non-critical $W_{2,s}(s=3,4)$ strings with two-spinor field
\cite{LiuYX NPB 2004}. The results show that there exists a
solution for the $W_{2,4}$ string but no solution for the
$W_{2,3}$ case. These results will be of importance for
constructing super $W$ strings, and they will provide the
essential ingredients.

However, all of these theories about the $W_{2,s}$ strings mentioned above are based on the non-linear $W_{2,s}$ algebras. Because of the intrinsic nonlinearity of the $W_{2,s}$ algebras, their study is a more difficult task compared to linear algebras. Fortunately, it has been shown that certain $W$ algebras can be linearized by the inclusion of a spin-1 current. This provides a way of obtaining new realizations of the $W_{2,s}$ algebras. Such new realizations were constructed for the purpose of building the corresponding scalar $W_{2,s}$ strings \cite{LuHong PLB 1995 351 179}. With the method, we have reconsidered the realizations of the non-critical $W_{3}$ string in Ref. \cite{zhang2005}, and found that there exists a solution of four-spinor field realization. In this paper, we will go on to construct the  nilpotent BRST charges of spinor non-critical $W_{2,4}$ string by using the linear bases of the $W_{1,2,4}$ algebra. Our results show that there exists a two-spinor field realization for non-critical $W_{2,4}$ string, which is just the same as the result in Ref. \cite{zhang2005}. All these results will be of importance for embedding of the Virasoro string into the $W_{2,4}$ string.

This paper is organized as follows. In section \ref{Non-criticalW24}, the method for constructing the BRST charge of non-critical $W_{2,4}$ string is given. In section \ref{W24Bases}, we first construct the linear bases of the  $W_{1,2,4}$ algebra with multi-spinor fields, then use these linear bases to construct the non-linear bases of the $W_{2,4}$ algebra and give the spinor realizations of them. Finally, a brief conclusion is given.

\section{BRST charge of non-critical $W_{2,4}$ string}\label{Non-criticalW24}

The non-critical $W_{2,4}$ string is the theories of $W_{2,4}$ gravity coupled to a matter system on which the $W_{2,4}$ algebra is realized. Now we give the spinor field realizations of them.

The BRST charge takes the form:
\begin{eqnarray}
Q_{B}&=&Q_{0}+Q_{1},\label{QB}\\
Q_{0}&=&\oint dz\; c(T^{eff}+ T_{\psi}+T_{M}+KT_{bc}+yT_{\beta\gamma}),\label{Q0}\\
Q_{1}&=&\oint dz\; \gamma F(\psi,\beta,\gamma,T_{M},W_{M}),\label{Q1}
\end{eqnarray}
where $K,y$ are pending constants and the operator $F(\psi,\beta,\gamma)$ has spin $s$ and ghost number zero. The energy-momentum tensors in (\ref{Q0}) are given by
\begin{eqnarray}
T_{\psi}&=&-\frac{1}{2}\partial\psi\psi,\label{Tpsi1}\\
T_{\beta\gamma}&=&4\beta\partial\gamma+3 \partial\beta\gamma,\label{Tbg1}\\
T_{bc}&=&2b\partial c+\partial bc,\label{Tbc1}\\
T^{eff}&=&-\frac{1}{2}\eta_{\mu\nu}\partial Y^{\mu} Y^{\nu},\label{Teff1}
\end{eqnarray}
and the matter currents $T_{M}$ and $W_{M}$, which have spin 2
and 4 respectively, generate the $W_{2,4}$ algebra.
The BRST charge generalizes the one for the scalar non-critical
$W_{2,4}$ string, and it is also graded with
$Q_{0}^{2}=Q_{1}^{2}=\{Q_{0},Q_{1}\}=0$. Again the first condition
is satisfied for $s=4$, and the remaining two conditions determine
the coefficients of the terms in
$F(\psi,\beta,\gamma,T_{M},W_{M})$ and $y$.

The most extensive combinations of $F$ in Eq. (\ref{Q1}) with correct spin and ghost number can be constructed as following:
\begin{eqnarray}\label{Fspin4}
F&=&   g_{1}\beta ^4\gamma ^4
     + g_{2}(\partial \beta)^2\gamma ^2
     + g_{3}\beta ^3\gamma ^2\partial \gamma
     + g_{4}\beta^2 (\partial \gamma )^2 \nonumber \\
&&
     + g_{5}\beta ^2\gamma ^2 \partial \psi \psi
     + g_{6}\partial \beta \gamma \partial \psi \psi
     + g_{7}\beta \partial \gamma \partial \psi \psi \nonumber \\
&&
     + g_{8}\beta  \partial ^2\beta \gamma ^2
     + g_{9}\partial ^2\beta \partial \gamma
     + g_{10}\partial \beta \partial ^2\gamma
     + g_{11}\partial ^2\psi \partial \psi  \nonumber \\
&&   + g_{12} \beta \partial ^3\gamma
     + g_{13}\partial ^3\psi \psi
     + g_{14}\partial \beta \gamma T_{M}
     + g_{15}\beta \partial \gamma T_{M} \nonumber \\
&&
     + g_{16}\partial\psi\psi T_{M}
     + g_{17}\beta\gamma\partial T_{M}
     + g_{18}T_{M}^2 \nonumber \\
&&
     + g_{19}\partial^2 T_{M}
     +g_{20}W_{M}.
\end{eqnarray}
Substituting (\ref{Fspin4}) back into Eq. (\ref{Q1}) and imposing the nilpotency conditions $Q_{1}^{2} = \{Q_{0},Q_{1}\}=0$, we can determine $y$ and $g_{i}(i=1,2,\cdots,20)$. They correspond to three sets of solutions, which had been given in Ref. \cite{LiuYX NPB 2004} and we don't list them here again.

In order to build the spinor non-critical $W_{2,4}$ string theory, we need the explicit construction for the matter currents $T_{M}$ and $W_{M}$. In Ref. \cite{LiuYX NPB 2004}, we had given the realization of them directly by the OPE relations of non-linear $W_{2,4}$ algebra.
In the following section, we will reconstruct the non-linear bases $T_{M}$ and $W_{M}$ of $W_{2,4}$ algebra by means of the linear bases $T_0$, $J_0$ and $W_0$ of $W_{1,2,4}$ algebra, making use of the fact that the $W_{2,4}$ algebra can be linearized through the addition
of a spin-1 current.

\section{Constructions of the non-linear bases of $W_{2,4}$ algebra}\label{W24Bases}

\subsection{Constructions of the linear bases of $W_{1,2,4}$ algebra}\label{W124}
We begin by reviewing the linearization of the $W_{2,4}$ algebra by the inclusion of a
spin-1 current \cite{Krivonos PLB 1994 335}. The linearized $W_{1,2,4}$ algebra take the
form:
\begin{eqnarray}\label{W123W124}
T_{0}(z)T_{0}(\omega) &\sim & \frac{C/2}{(z - w)^4}+\frac{2T_{0}(\omega)}{(z-\omega)^2}+
\frac{\partial
T_{0}(\omega)}{z-\omega}, \nonumber \\
 T_{0}(z)W_{0}(\omega) &\sim &
\frac{4 W_{0}(\omega)}{(z-\omega)^2}+ \frac{\partial
W_{0}(\omega)}{z-\omega}, \nonumber \\
T_{0}(z)J_{0}(\omega) &\sim &  \frac{C_{1}}{(z -
w)^3}+\frac{J_{0}(\omega)}{(z-\omega)^2}+ \frac{\partial
J_{0}(\omega)}{z-\omega}, \\
J_{0}(z)J_{0}(\omega) &\sim & \frac{-1}{(z-\omega)^2},\nonumber \\
J_{0}(z)W_{0}(\omega) &\sim & \frac{hW_{0}(\omega)}{z-\omega},~~
W_{0}(z)W_{0}(\omega) \sim  0,\nonumber
\end{eqnarray}
where the coefficients $C$, $C_{1}$ and  $h$ are given by
\begin{eqnarray}\label{CC1h}
  C =86+30t^{2}+\frac{60}{t^{2}},\;\;\;
   C_{1}=-3t-\frac{4}{t},\;\;\;
   h=t.
\end{eqnarray}

To obtain the realizations for the linearized $W_{1, 2, 4}$ algebra, we use the multi-spinor fields $\psi^{\mu}$, which have spin 1/2 and satisfy the OPE
\begin{equation}
\psi^{\mu}(z)\psi^{\nu}(\omega) \sim - \frac{1}{z-\omega} \delta^{\mu\nu},
\end{equation}
to construct the linear bases of them. The general forms of these linear bases can be
taken as follows:
\begin{equation}
T_0= -\frac{1}{2} \partial \psi^{\mu} \psi^{\mu},~
J_0 = \alpha_{\mu \nu} \psi^{\mu} \psi^{\nu}(\mu < \nu),~
W_0=0,
\end{equation}
where $\alpha_{\mu \nu}$ are pending coefficients. By making use of the OPE
$J_{0}(z)J_{0}(\omega)$ in (\ref{W123W124}), we can get the equation that the
coefficients $\alpha _{\mu \nu}$ satisfy, i.e. $\sum _{\mu < \nu} \alpha _{\mu \nu}^2
=1$. From the OPE relation of $T_{0}$ and $J_{0}$, it is easy to get $C_1=0$.
Substituting the value of $C_1$ into Eq. (\ref{CC1h}), we can get the value of $t$. Then
the total central charge $C$ for $T_0$ can be obtained from Eq. (\ref{CC1h}). So we can
determine the explicit form of $T_{0}$ under the restricted condition of its central
charge. Finally, using the OPE $T_{0}(z)J_{0}(\omega)$ in (\ref{W123W124}) again, we get
the exact form of $J_{0}$. The complete results are listed as follows:
\begin{eqnarray}
t&=& \pm \frac{2}{\sqrt{3}}i,~ C=1,~ C_{1}=0, \label{tCC1} \\
T_0&=&-\frac{1}{2}\sum_{\mu=1}^{2}\partial\psi^{\mu}\psi^{\mu},~ J_0=\pm\psi^{1}\psi^{2},~
W_0=0.\label{T0J0W0spin4}
\end{eqnarray}

\subsection{Spinor realizations of the $W_{2,4}$ algebra
based on the linear $W_{1,2,4}$ algebra}\label{W24algebra}

Now let us consider the spinor realizations of the non-linear $W_{2,4}$ algebra with linear bases of the $W_{1,2,4}$ algebra.
In conformal OPE language the $W_{2,s}$ algebra takes the following forms:
\begin{eqnarray}
T(z)T(\omega) &\sim& \frac{C/2}{(z-w)^4}%
                     +\frac{2T(\omega)}{(z-\omega)^2}%
                     +\frac{{\partial}T(\omega)}{(z-\omega)},\\
T(z)W(\omega) &\sim& \frac{s W(\omega)}{(z-\omega)^2}%
                     +\frac{{\partial}W(\omega)}{(z-\omega)},\\
W(z)W(\omega) &\sim& \frac{C/s}{(z-w)^{2s}} %
                     +\sum_{\alpha}\frac{P_{\alpha}(\omega)}%
                                        {(z-\omega)^{\alpha+1}},
\end{eqnarray}
in which $P_{\alpha}(\omega)$ are polynomials in the primary fields $W$, $T$ and their
derivatives. For an exact $s$, the precise form of $W$ and the corresponding central
charge $C$ can be solved by means of these OPEs. The OPE $W(z)W(\omega)$ for $W_{2,4}$ algebra
takes the form \cite{Kausch NPB 1991}:
\begin{eqnarray}\label{OPEofWWspin4}
W(z)W(\omega) &\sim&
   \frac{C/4}{(z-\omega)^8}
   +\frac{2 T}{(z - w)^6}
   +\frac{\partial T}{(z-\omega)^5}\nonumber \\
&+& \frac{3}{10}\frac{\partial ^2T}{(z-\omega)^4}
   +\frac{b_{1} U}{(z-\omega)^4}
   +\frac{b_{2} W}{(z-\omega)^4} \nonumber \\
&+& \frac{1}{15}\frac{\partial ^3 T}{(z-\omega)^3}
   +\frac{1}{2} \frac{b_{1}\partial U}{(z-\omega)^3}
   +\frac{1}{2} \frac{b_{2}\partial W}{(z-\omega)^3}\nonumber \\
&+& \frac{1}{84}\frac{\partial ^4 T}{(z-\omega)^2}
   +\frac{5}{36} \frac{b_{1}\partial ^2 U}{(z-\omega)^2}
   +\frac{5}{36} \frac{b_{2}\partial^2 W}{(z-\omega)^2}
    \nonumber\\
&+&\frac{b_{3} G}{(z-\omega)^2}
   +\frac{b_{4}A}{(z-\omega)^2}
   +\frac{b_{5}B}{(z-\omega)^2} \nonumber \\
&+& \frac{1}{560}\frac{\partial^5T}{(z-\omega)}
   +\frac{1}{36} \frac{b_{1}\partial ^3 U}{(z-\omega)}
   +\frac{1}{36} \frac{b_{2}\partial^3 W}{(z-\omega)}
     \nonumber\\
&+& \frac{1}{2}\frac{b_{3}\partial G}{(z-\omega)}
   +\frac{1}{2}\frac{b_{4}\partial A}{(z-\omega)}
   +\frac{1}{2} \frac{b_{5}\partial B}{(z-\omega)},
\end{eqnarray}
where the composites $U$ (spin 4), and $G$, $A$ and $B$ (all spin 6), are defined by
\begin{eqnarray}
U &=& (TT)-\frac{3}{10} \partial ^2 T,\nonumber\\
G &=& (\partial ^2 T T)-\partial({\partial}TT)%
      +\frac{2}{9}{\partial}^{2}(TT)%
      -\frac{1}{42}{\partial}^{4}T, \nonumber\\
A &=& (TU)-\frac{1}{6}\partial ^2 U,~~~
B = (TW)-\frac{1}{6}\partial ^2 W,\nonumber
\end{eqnarray}
with normal ordering of products of currents understood. The coefficients
$b_{1},b_{2},b_{3},b_{4}$ and $b_{5}$ are given by
\begin{eqnarray}
b_{1} &=& \frac{42}{5C+22},\nonumber\\
b_{2} &=& \sqrt{\frac{54(C+24)(C^2-172C+196)}%
                     {(5C+22)(7C+68)(2C-1)}},\nonumber\\
b_{3} &=& \frac{3(19C-524)}{10(7C+68)(2C-1)},\nonumber\\
b_{4} &=& \frac{24(72C+13)}{(5C+22)(7C+68)(2C-1)},\nonumber\\
b_{5} &=& \frac{28}{3(C+24)}b_{2}. \nonumber
\end{eqnarray}

The bases of the $W_{2,4}$ algebra can be constructed by the linear bases of the
$W_{1,2,4}$ algebra:
\begin{eqnarray}
T = T_{0}, ~~~
W = W_{0}+W_{R}(J_{0},T_{0}),
\end{eqnarray}
where the currents $T_{0}$, $W_{0}$ and $J_{0}$ generate the $W_{1,2,4}$ algebra and have
been constructed with multi-spinor fields in Eq. (\ref{T0J0W0spin4}). First we can write down the most
general possible structure of $W$. Then the relations of above OPEs of  $T$ and $W$
determine the coefficients of the terms in $W$. Finally, substituting these coefficients
and $T_{0}$, $J_{0}$ and $W_{0}$ which have been constructed in section II into the
expressions of $T$ and $W$, we can get the spinor realizations of the $W_{2,4}$ algebra.
The explicit results turn out to be very simple as follows:
\begin{eqnarray}
T &=& T_0, \label{Tspin4_1}\\
W &=& W_0 + \eta_{1}\partial^{3}J_0%
      +\eta_{2}\partial^{2}J_0 J_0%
      +\eta_{3}({\partial}J_0)^2  \nonumber\\
&&
      +\eta_{4} \partial J_0 (J_0)^2
      +\eta_{5}(J_0)^4
      +\eta_{6} \partial ^2 T_0%
      +\eta_{7}(T_0)^2   \nonumber\\
&&
      +\eta_{8} \partial T_0 J_0
      +\eta_{9}T_0 \partial J_0%
      +\eta_{10}T_0(J_0)^2, \label{Wspin4_1}
\end{eqnarray}
where
\begin{eqnarray}
\eta_{1} &=& \eta_{0} (1800 + 5562 t^2 + 7744 t^4 + 6167 t^6 \nonumber\\
         &&   + 2631 t^8 + 450 t^{10}),\nonumber\\
\eta_{2} &=& 12 \eta_{0} t (450 + 1278 t^2 + 1429 t^4%
            + 752 t^6 + 150 t^8),\nonumber \\
\eta_{3} &=& 6 \eta_{0} t (1050 + 2932 t^2 + 3009 t^4%
            + 1353t^6 + 225 t^8),\nonumber \\
\eta_{4} &=& 12 \eta_{0} t^2 (4 + 3 t^2)%
            (150 + 226 t^2 + 75 t^4),\nonumber \\
\eta_{5} &=& 6 \eta_{0} t^3 (150 + 226 t^2 + 75 t^4),\nonumber \\
\eta_{6} &=& 3 \eta_{0} t (240 + 724 t^2 + 865 t^4%
                           + 465 t^6 + 90 t^8),\nonumber \\
\eta_{7} &=& 6 \eta_{0} t^3 (3 + t^2)(32 + 27 t^2),\nonumber \\
\eta_{8} &=& 12 \eta_{0} t^2 (150 + 376 t^2 + 301 t^4 %
                              + 75 t^6),\nonumber \\
\eta_{9} &=& 12 \eta_{0} t^2 (300 + 602 t^2 + 376 t^4 %
                              + 75 t^6),\nonumber \\
\eta_{10} &=& 12 \eta_{0} t^3 (150 + 226 t^2 + 75 t^4),\nonumber \\
\eta_{0} &=& -(36t+12t^3)^{-1}(504000 + 3037560t^2
              + 7617488 t^4  \nonumber \\
&&
             + 10300470 t^6 + 8109196 t^8 + 3716751 t^{10}\nonumber \\
&&
             + 918585 t^{12}%
             + 94500 t^{14})^{-1/2}.\nonumber
\end{eqnarray}
Making use of the values of $t$ in Eq. (\ref{tCC1}), we can get the simple form of $W$
as follows:
\begin{eqnarray}\label{Wspin4_2}
W\!&=&\!   W_0 %
      + \frac{\sqrt{2}}{360}%
        \left(\mp 54 i \partial ^3 J_0 %
              + 198 \sqrt{3}\partial ^2 J_0 J_0
              + 27 \sqrt{3}({\partial}J_0)^2 \right.\nonumber \\
  &&
              - 108 \sqrt{3}(J_0)^4
              + 12\sqrt{3} \partial ^2 T_0
              - 40\sqrt{3}(T_0)^2 \nonumber \\
  &&         \left. \mp 108i \partial T_0 J_0 %
              \pm 216i T_0 \partial J_0
             -216 \sqrt{3}T_0 (J_0)^2 \right).
\end{eqnarray}
Substituting the expressions of $T_{0}$, $J_{0}$ and $W_{0}$ in
Eq. (\ref{T0J0W0spin4}) into Eqs. (\ref{Tspin4_1}) and
(\ref{Wspin4_2}), we immediately obtain the explicit constructions
of $T$ and $W$ for the $W_{2,4}$ algebra as follows:
\begin{eqnarray}
T &=& -\frac{1}{2}%
      \left(\partial \psi^{1}\psi^{1}%
            +\partial \psi^{2}\psi^{2}\right), \label{Tspin4}\\
W &=& \frac{1}{72\sqrt{6}} %
            \left(\partial^3\psi^{1}\psi^{1} %
                  -9\partial^2\psi^{1}\partial\psi^{1}%
                  +84\partial\psi^{1}\psi^{1}%
                   \partial\psi^{2}\psi^{2}\right.\nonumber\\
  &&             +\left.\partial^3\psi^{2}\psi^{2}%
                  -9\partial^2\psi^{2}%
                    \partial\psi^{2}\right).\label{Wspin4}
\end{eqnarray}
This result is just the same as that given in Ref. \cite{LiuYX NPB 2004}. But the result here is obtained from the method based on linearized algebra. Substituting the form (\ref{Fspin4}) of $F$ into (\ref{Q1}), with $T_{M}$ and $W_{M}$ given by Eqs. (\ref{Tspin4}) and (\ref{Wspin4}) respectively, we get the exact spinor field realizations of the non-critical $W_{2,4}$ string.

\section{Conclusion}\label{ConclusionW2s}

In this paper, we have reconstructed the spinor field realizations
of the non-critical $W_{2,4}$ string, making use of the fact that
the $W_{2,4}$ algebra can be linearized through the addition of a
spin-1 current. First, the construction for the BRST charge of
non-critical $W_{2,4}$ string is given. Subsequently, we use the
multi-spinor fields $\psi^{\mu}$ to construct the linear bases of
the linearized $W_{1,2,4}$ algebra. Finally, the non-linear bases
of $W_{2,4}$ algebra is constructed with these linear bases $T_0,
J_0$ and $W_0$. Thus, the spinor field realizations of the
non-critical $W_{2,4}$ string are obtained, and the result is the
same as that of our previous work \cite{LiuYX NPB 2004}. It is
worth pointing out that the fields which give the realizations of
the non-critical $W_{2,4}$ string are two-spinor, while the fields
corresponding to the non-critical $W_{2,3}$ string are
four-spinor. The reason is that, the total central charges $C$ of
$T_0$ for $W_{1,2,3}$ algebra and $W_{1,2,4}$ algebra are 2 and 1,
respectively, and the central charge for any term $-(1/2)\partial
\psi^{\mu} \psi^{\mu}$ in $T_0$ is 1/2. Compared with the results
of Ref. \cite{LiuYX JHEP 2005}, in which the BRST charge of
non-critical $W_{2,4}$ string is constructed by ghost fields, the
realizations of spinor fields in this paper are more simple. It is
very clear that the spinor realizations of the critical $W_{2,4}$
string can be obtained when the matter currents $T_{M}$ and
$W_{M}$ in Eqs. (\ref{Q0}) and (\ref{Q1}) vanish, and the
corresponding constructions become relatively simple. We expect
that there should exist such realizations for the case of higher
spin $s$.

\section*{Acknowledgements}
It is a pleasure to thank Prof. Y.S. Duan and Dr. H. Wei for useful discussions. We have
also made extensive use of a Mathematica package for calculating OPEs, written by Prof.
K. Thielemans \cite{Thielemans_IJMP_1991}.

\end{document}